# POPULATION SYNTHESIS OF X-RAY SOURCES AT THE GALACTIC CENTER

V.M. Lipunov[1], L.M. Ozernoy[2], S.B. Popov[1], K.A. Postnov[1], M.E. Prokhorov[1]

[1]Sternberg Astronomical Institute, Universitetskij pr. 13, 119899 Moscow

[2]Physics Dept. and Inst. for Comput. Sci. & Inform., George Mason Univ.,
also Lab. for Astron. & Solar Phys., NASA/GSFC

## Abstract

X-ray binary population at the Galactic center is proposed here to form as a result of a starburst, which is synthesized by using the "Scenario Machine" code of binary star evolution. For the currently assumed starburst age of 4-7 millions years, our results are consistent with the recent *GRANAT* X-ray observations of the Galactic center and predict a substantial number of X-ray binaries with black holes.

# 1  Introduction

X-ray observations of the Galactic center have revealed a number of energetic X-ray sources to be located in the innermost regions of the Galaxy (Pavlinsky et al[9], Churasov et al[2] ), with a part of them being attributed, by their spectral characteristics, to black hole (BH) candidates. These observations demonstrate an enhanced, compared to the average galactic value, spatial density of X-ray binary systems in the central region of the Galaxy. In the present paper, we show that such situation is a natural consequence of binary stellar evolution if a star formation burst occurred a few millions years ago at the Galactic center. The most recent evidence for such starburst was suggested by Sofue[10] using the North Polar Spur data and by Ozernoy[8] using the data on the 10 KeV gas in the cental 200 pc of the Galaxy. Also, recently, Blum et al[1] discovered WR-star in the Galaxtic center. This is a very strong argument for the recent star formation burst in the Galactic center. Here, we apply Monte Carlo simulations to the evolution of a large ensemble of binary systems, with proper accounting for spin evolution of magnetized compact stars such as white dwarfs (WDs) and neutron stars (NSs) (Lipunov[4], Lipunov et al.[7] ). This method has been shown to be a powerful tool for studying different products of stellar evolution both in spiral and elliptical galaxies under a wide range of initial conditions and star formation histories (see e.g. Kornilov & Lipunov[3], Lipunov[6], Lipunov et al.[7] and references therein). In this paper, we will focus on the most prominent, from the observational point of view, representatives of the late stages of massive binary evolution, such as X-ray transients (NS in a highly eccentric orbit around a main sequence star, like A0535+26), super-accreting BHs (observationally seen as SS 433 if in pair with a Roche lobe filling secondary component), and BH + supergiant binaries (like Cyg X-1, with an evolved supergiant underfilling its Roche lobe).

# 2  The model

Suppose that an instantaneous burst of star formation occured at the Galactic center, with a half stars being in binaries. As we are interested in the evolution of NS and BH in binaries, we consider only massive binaries, i.e. those having the primary component mass in the range of 10-120 $M_\odot$ and distributed according to the Salpeter's mass function with an power law exponent $\alpha = 2.35$. The initial mass ratio distribution (which is as yet a controversial issue) was taken in a power law form $f(q) \propto q^2$ strongly favouring binaries with equal masses (Lipunov et al 1995, in preparation).

The total mass of stars formed during the starburst was about $4 \times 10^5 M_\odot$ (Tamblyn & Rieke[11] ). To get statistically significant results, the evoluiton of 300,000 binary systems have been computed. Then we normalized the figures so as to be in agreement with the Tamblyn & Rieke[11] calculations of the number of massive OB-stars that survive $\sim 7$ Myr after the starburst.

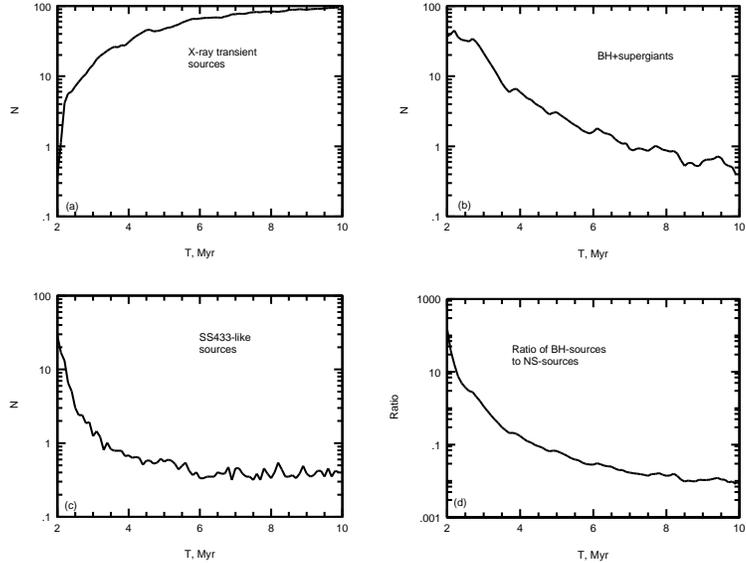

Figure 1: The evolution of some selected types of X-ray binaries during the starburst at the Galactic center.

# 3   The results

The evolution of some selected types of X-ray binaries during the first 10 million years after the onset of star formation burst is presented in Figure 1.

First we show (Fig. 1a) the number of X-ray transient sources consisting of a NS in an eccentric orbit around a massive secondary that acquired enough angular momentum during the mass exchange stage to become a rapidly rotating Be star. To become an X-ray transient, the NS must accrete matter from the secondary Be-star, at least during the periastron passages. In fact, not all the transients are in the accretion stage at the same moment, so that the observed number of these sources can be a few times less.

Then we show the evolution of a BH containing X-ray binaries of Cyg X-1 type (Fig. 1b) and of those with a superaccreting BH of SS 433 type (Fig. 1c). We note that the well known X-ray source 1E 1740-2942 (Churasov et al[2]) may belong to SS 433-like binaries.

One of the most interesting results concerns the relative number of SS 433-like binaries together with Cyg X-1–like binaries and X-ray transient sources. Our calculations yield for the ratio of these classes of sources to be $\sim (1/5 - -1 : 40)$, which significantly exceeds this ratio for the entire Galaxy.

The number ratio of a BH containing binaries (of both SS 433 and Cyg X-1 type) to the X-ray transients with Be-stars is plotted in Fig. 1d. It can be seen that this ratio is very sensitive to the time elapsed after the starburst and therefore it can serve as a tool to estimate the age of X-ray binaries at the Galactic center.

# 4  Discussion and conclusions

We compare our results with the real data by using the *GRANAT* X-ray observations of the Galactic center (Churazov et al[2], Pavlinsky et al [9]). 11 X-ray sources have been reported to be observed in the central region of the Galaxy (750 pc × 750 pc), including 2 sources being classified as BH-candidates by their spectra, and 9 being X-ray transients.

As we noted above, the BH-candidates/X-ray transients ratio is a good indicator of the time passed after the onset of the starburst. The computed ratio$\approx 0.2$ at the age of 4 Myr and $\approx 0.02$ at the age of 7 Myr (which should be considered as a lower limit to the true value because we are not able to observe all the X-ray transients simultaneously) roughly corresponds to the observed ratio of $\approx 0.2$. Furthermore, in our model at the age of 7 Myr the X-ray transients occupy a region $\approx$ 2000 pc × 2000 pc in size (but more than the half of them are distributed in the region $\approx$ 700 pc × 700 pc), as they acquire high velocities (after the faster evolving primary component explodes as a supernova) and move with an initially high velocity dispersion of about 100 km/s. Hence, by considering the central 750 pc × 750 pc region, one will get the ratio closer to the observed value.

To conclude, we note that the modelling of standard binary stellar evolution after a $\sim 4 \times 10^5 M_\odot$ starburst at the Galactic center about 4-7 million years ago yields the number of X-ray sources to be consistent with available X-ray observations of the Galactic center.

*Acknowledgements*. The authors acknowledge I. Panchenko and S. Nazin for useful discussions. The work was partially supported by the COSMION grant and by Grant No JAP100 from the International Science Foundation and Russian Government.